# Interferometric Motion Detection in Atomic Layer 2D Nanostructures: Visualizing Signal Transduction Efficiency and Optimization Pathways

Zenghui Wang, Philip X.-L. Feng[*]

*Department of Electrical Engineering & Computer Science,*

*Case School of Engineering, Case Western Reserve University,*

*10900 Euclid Avenue, Cleveland, OH 44106, USA*

**Atomic layer crystals are emerging building blocks for enabling new two-dimensional (2D) nanomechanical systems, whose motions can be coupled to other attractive physical properties in such 2D systems.  Optical interferometry has been very effective in reading out the infinitesimal motions of these 2D structures and spatially resolving different modes. To quantitatively understand the detection efficiency and its dependence on the device parameters and interferometric conditions, here we present a systematic study of the intrinsic motion responsivity in 2D nanomechanical systems using a Fresnel-law-based model.  We find that in monolayer to 14-layer structures, MoS$_2$ offers the highest responsivity among graphene, h-BN, and MoS$_2$ devices and for the three commonly used visible laser wavelengths (633, 532, and 405nm).  We also find that the vacuum gap resulting from the widely used 300nm-oxide substrate in making 2D devices, fortunately, leads to close-to-optimal responsivity for a wide range of 2D flakes.  Our results elucidate and graphically visualize the dependence of motion transduction responsivity upon 2D material type and number of layers, vacuum gap, oxide thickness, and detecting wavelength, thus providing design guidelines for constructing 2D nanomechanical systems with optimal optical motion readout.**

[*]Corresponding Author.  Email: philip.feng@case.edu

Atomic layer crystals such as graphene, hexagonal boron nitride (h-BN), and transition metal dichalcogenides (TMDCs) have emerged as a new class of two-dimensional (2D) materials, exhibiting promises for both fundamental research and technological applications[1,2,3,4,5]. Amongst their many attributes, the excellent mechanical properties (*e.g.*, large elastic moduli, ultrahigh mechanical strength, and superb strain limits[6] up to 20% to 30%) make these materials attractive for constructing 2D nanoelectromechanical systems (NEMS)[7], providing opportunities for coupling their material properties across multiple information-transduction domains (*e.g.*, electrical, optical, mechanical), down to individual atomic layers. Analogous to simple harmonic oscillators being essential to mechanical systems and dynamics at any scale, 2D nanomechanical resonators are particularly interesting prototypes of 2D NEMS[8,9,10,11]. While various electrical[9,12,13,14,15], mechanical[16], and optical[10,11,13,17,18,19] motional signal transduction schemes have been employed for 2D harmonic resonators, laser optical interferometry clearly stands out as a very important and widely used technique.

Interferometric motion detection has been playing an important role in scientific explorations since before the 1900s[20]; today it is used in many research fields to detect various physical quantities—*e.g.*, from gravitational waves[21] to single electron spins[22]—with the state-of-the-art achieving fractional displacement (*i.e.*, 'strain') sensitivity of $10^{-23}/\sqrt{\text{Hz}}$ (in $10^2$–$10^3$Hz), defying the quantum shot-noise limit[23,24]. Laser interferometry has also played a critical role in NEMS research to demonstrate motion detection of NEMS resonators with a variety of geometries, such as cantilevers[25,26], beams[27,28], and wires[29,30]. As such 1D nanostructures continue to scale toward the atomic level, it becomes more challenging to detect their motions interferometrically[28,31] because the light intensity reflected from the molecular-scale nanostructure (*e.g.*, single-wall carbon nanotube) diminishes. In contrast, 2D NEMS structures are intrinsically compatible with interferometric detection scheme[32], as the added dimension ensures sufficient light reflection from the planar surfaces. Among the different motion detection schemes applicable to 2D NEMS structures, optical interferometry exhibits important advantages. First, optical interferometry is suitable for any 2D material, and imposes minimal requirement of device geometry (*e.g.*, it does not require electrodes or conductivity)—as long as the device moves, its motion may be detected interferometrically[32]. More importantly, laser interferometry boasts excellent motion detection sensitivity (down to fm/Hz$^{1/2}$ level[32]): to date it is the only technique capable of measuring 2D NEMS' completely undriven thermomechanical motions that set the fundamental limit for the smallest detectable motion at given temperature[10,11,17]. The very first resonances of graphene, TMDC, and black phosphorus 2D NEMS have all been detected by using optical interferometric techniques[10,11,17].

While interferometric resonance detection in 2D NEMS has been demonstrated, the fundamental effects in optical signal transduction of 2D device motions, limits, and scaling laws remain to be systematically investigated. Moreover, thanks to their new optical properties, 2D atomic layer optical interferometric systems are distinct from the ones involving conventional MEMS/NEMS[25,26,27,29,32] (*e.g.*, in Si, SiN, SiC, and AlN materials; in structures such as cantilevers, beams, membranes and disks), demanding a dedicated investigation. Specifically, device parameters such as 2D material type, number of atomic layers, vacuum gap, and detecting wavelength, can all affect the motion responsivity (the transduction efficiency). This is analogous, to certain extent, to the fact that oxide thickness and illuminating wavelength affect the optical visibility of the 2D material on substrate[33,34,35,36]. Here, we present a systematic study of the motion responsivity. Using a Fresnel-law optical model, we quantitatively illustrate and visualize how the different device parameters affect the motion responsivity, and delineate pathways towards device structures with optimized motion transduction efficiencies for various 2D materials.

**Reflectance and Responsivity.** Figure 1 illustrates the device structure in our model. As in most experiments, we consider normal laser incidence from vacuum onto this tetra-layer structure composed of 2D crystal (subscript "2D" in equations), the vacuum gap, the remaining SiO$_2$ layer, and the Si wafer (treated as semi-infinite medium). The reflectance $R$ of the structure (total fraction of light reflected) is determined by the interference of the reflected light from all interfaces. Analysis of the multiple reflections inside this tetra-layer structure gives[33,37]:

$$R = \left| \frac{\boldsymbol{r_1} + \boldsymbol{r_2}e^{-2i(\varphi_1)} + \boldsymbol{r_3}e^{-2i(\varphi_1+\varphi_2)} + \boldsymbol{r_4}e^{-2i(\varphi_1+\varphi_2+\varphi_3)} + \boldsymbol{r_1r_2r_3}e^{-2i(\varphi_2)} + \boldsymbol{r_1r_3r_4}e^{-2i(\varphi_3)} + \boldsymbol{r_1r_2r_4}e^{-2i(\varphi_2+\varphi_3)} + \boldsymbol{r_2r_3r_4}e^{-2i(\varphi_1+\varphi_3)}}{1 + \boldsymbol{r_1r_2}e^{-2i(\varphi_1)} + \boldsymbol{r_1r_3}e^{-2i(\varphi_1+\varphi_2)} + \boldsymbol{r_1r_4}e^{-2i(\varphi_1+\varphi_2+\varphi_3)} + \boldsymbol{r_2r_3}e^{-2i(\varphi_2)} + \boldsymbol{r_3r_4}e^{-2i(\varphi_3)} + \boldsymbol{r_2r_4}e^{-2i(\varphi_2+\varphi_3)} + \boldsymbol{r_1r_2r_3r_4}e^{-2i(\varphi_1+\varphi_3)}} \right|^2$$

(1)

where $\boldsymbol{r_1}$ through $\boldsymbol{r_4}$ are the reflection coefficients at the vacuum-2D, 2D-vacuum, vacuum-SiO$_2$, and SiO$_2$-Si interfaces, respectively, and $\boldsymbol{\varphi_1}$, $\boldsymbol{\varphi_2}$, and $\boldsymbol{\varphi_3}$ are the corresponding phase changes:

$$\boldsymbol{r_1} = \frac{n_{vacuum} - \boldsymbol{n_{2D}}}{n_{vacuum} + \boldsymbol{n_{2D}}}, \quad \boldsymbol{r_2} = \frac{\boldsymbol{n_{2D}} - n_{vacuum}}{\boldsymbol{n_{2D}} + n_{vacuum}}, \quad \boldsymbol{r_3} = \frac{n_{vacuum} - \boldsymbol{n_{SiO_2}}}{n_{vacuum} + \boldsymbol{n_{SiO_2}}}, \boldsymbol{r_4} = \frac{\boldsymbol{n_{SiO_2}} - \boldsymbol{n_{Si}}}{\boldsymbol{n_{SiO_2}} + \boldsymbol{n_{Si}}},$$

(2)

$$\boldsymbol{\varphi_1} = \frac{2\pi \boldsymbol{n_{2D}} d_1}{\lambda}, \quad \boldsymbol{\varphi_2} = \frac{2\pi n_{vacuum} d_2}{\lambda}, \quad \boldsymbol{\varphi_3} = \frac{2\pi \boldsymbol{n_{SiO_2}} d_3}{\lambda},$$

(3)

where $d_1$ is the 2D crystal thickness, $d_2$ is the vacuum gap depth, $d_3$ is the SiO$_2$ thickness, and $\lambda$ is laser wavelength. Note that bold fonts indicate complex variables.

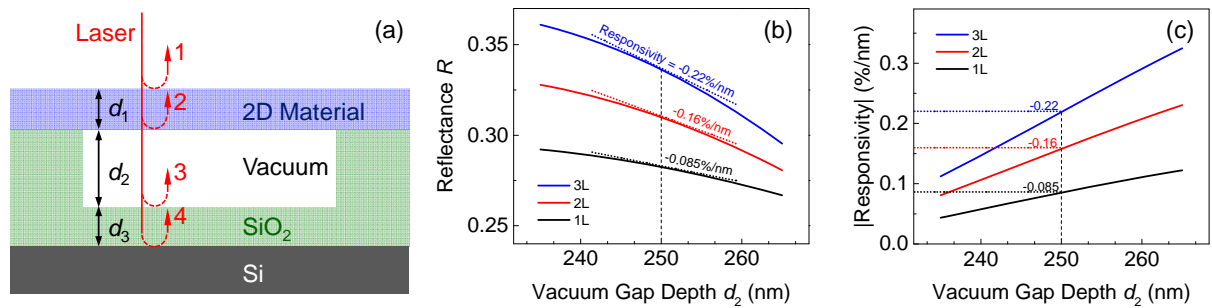

**Figure 1**: Interferometry detection of 2D device motion. (a) The device model and illustration of 1$^{st}$-order reflections. (b) Reflectance $R$ *vs.* vacuum gap depth $d_2$ for 1-3L MoS$_2$ devices. Dashed line indicates the example geometry ($d_2$=250nm), and the slopes of the $R$-$d_2$ curves gives the responsivity $\Re$ (dotted lines). (c) Magnitude of motion-to-reflectance responsivity for 1L, 2L, and 3L MoS$_2$ devices. Dotted horizontal lines show the values of $|\Re|$ at $d_2$=250nm (vertical dashed line).

As the 2D layer assumes flexural motion, the vacuum gap depth $d_2$ changes, leading to modification of the reflectance $R$ (Eq. 1). Here we use $MoS_2$ as an example to illustrate this signal transduction process, as the real part of its index of refraction is much higher than other 2D materials (see Methods), resulting in much stronger reflection from the crystal surfaces ($r_1$ and $r_2$ in Eq. 2) and consequently stronger interferometric motion transduction, particularly important for ultrathin samples which transmit most of the light. Figure 1b shows the reflectance $R$ (at 633nm laser illumination) as a function of $d_2$ for mono-, bi-, and trilayer (1L, 2L, and 3L) $MoS_2$ devices with $d_2$=250nm and $d_3$=50nm, showing that $R$ varies smoothly with $d_2$, and increase with number of layers (or thickness $d_1$) over this range of $d_2$.

The photodetector measures motion-induced changes in reflectance. Therefore, the greater change in $R$ per unit device motion, the more efficient the signal transduction. This motion-to-reflectance responsivity $\Re$ is defined as $\Re = \partial R / \partial d_2$: thus the slope of $R(d_2)$ at $d_2$ =250nm (Fig. 1b) represents the values of $\Re$ in these devices. Note that $\Re$ can be negative (and its magnitude, not sign, determines the effectiveness in detecting device motion), and in Fig. 1c the magnitude $|\Re|$ is plotted. It can be seen that in this range, $\Re$ increases roughly linearly with device layer number and $d_2$.

**Visibility *vs*. Responsivity.** The 'visibility'[33,35,36] and motion responsivity in 2D materials are related but different. While rooted from the same formalism (Eq. 1), they focus on different aspects of the equation. For visibility, a larger optical contrast (greater change in $R$) between two locations: one with 2D crystal ($d_1 \neq 0$) and one without ($d_1$=0) gives better visibility; whereas for motion detection, a greater responsivity (steeper slope in $R$ *vs*. $d_2$) leads to greater signal given the same motion amplitude. The optimal detection scheme is also different: when searching for 2D flakes in microscope, as the human eye is highly color-sensitive, it is desirable to use a multi-color (such as white light) illumination, and when $R$ increases for one color and decreases for another it gives enhanced color contrast. In motion detection, the photodetector measures the change in light intensity, and using a monochromatic light source is more desirable as it removes the possible cancellation between different wavelengths (*i.e.*, $R$ increases for one $\lambda$ and decreases for another). Therefore, instead of calculating over a range of continuous $\lambda$ values (as in works focusing on visibility of on-substrate materials[35]), we focus on monochromatic illumination, mostly $\lambda$=633, 532, and 405nm, which are widely used in 2D structures motion detection.

Figure 2 shows the calculated motion responsivity $\Re$ (633nm illumination) of 1L, 2L, and 3L $MoS_2$ over a large range of $d_2$ values, for devices fabricated on 300nm $SiO_2$ substrate followed by oxide etch (thus $d_2+d_3$=300nm). Note that over large range of $d_2$, $\Re$ can cross 0 (where the motions lead to no reflectance variation) and change sign.

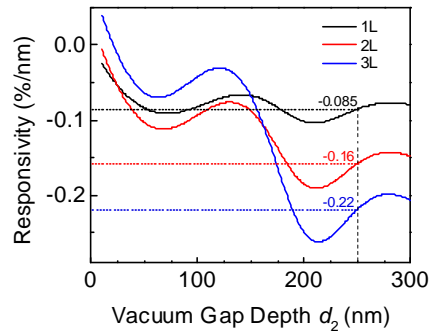

**Figure 2**. Responsivity $\Re$ of 1L, 2L, and 3L $MoS_2$ device for a larger $d_2$ range (10 to 300nm). Vertical dashed line ($d_2$=250nm) allows easy comparison with Fig. 1, with dotted horizontal lines showing the values of $\Re$.

**Dependence on Device Structure.** Recently emerging transfer techniques[38] make it possible to fabricate suspended 2D devices on arbitrary substrate structures. Therefore, in analysis below we vary $d_1$, $d_2$, and $d_3$ independently. Figure 3a illustrate the results for 1L, 2L and 3L $MoS_2$ devices. We make the following observations. First, in this comprehensive parameter space (both $d_2$ and $d_3$ covering multiples of $\lambda$), periodic behavior is evident. Second, these devices exhibit similar locations ($d_2$ $d_3$ combinations) for the $\Re$ peaks (both positive and negative). Third, the magnitude of $\Re$ (and thus the amplitude of $\Re$ variation) increases with number of layers for 1-3L devices.

We first focus on the $|\Re|$ peak locations by using 2D color plots (Fig. 3b,c). We find the same $\Re$ periodicity (along both $d_2$ and $d_3$) for all device thicknesses. In $d_2$ direction, the period is exactly $\lambda/2$ ($\lambda$=633nm). This is because when $d_2$ changes by $\lambda/2$, the total optical path for rays 3&4 (Fig. 1a) changes exactly by $\lambda$, keeping the interferometric condition unchanged. The periodicity along the $d_3$ axis is $\lambda/2\,\boldsymbol{n}_{SiO_2}$, which can be understood in a similar pattern by considering the optical path inside $SiO_2$. Note that the 0 optical absorption of $SiO_2$ at 633nm[39] helps ensure the perfect repeatability in the $d_3$ direction (*i.e.*, including both periodicity in $d_3$ and identical amplitude along this periodic pattern).

We thus plot gridlines at $d_2$=$n\lambda/2$ and $d_3$=$m\lambda/2\,\boldsymbol{n}_{SiO_2}$ ($n$, $m$ are integers) to better visualize the periodicity. We find that all the $|\Re|$ peaks are located along $d_3$=$m\lambda/2\,\boldsymbol{n}_{SiO_2}$. In contrast, their $d_2$ values gradually vary with $d_1$, more apparent in the positive peaks (dark red). By compiling the results for 1L to 50L $MoS_2$ devices into a 3D stack (Fig. 4a), with device thickness $d_1$ in the third (vertical) dimension, we confirm that $d_3$=$m\lambda/2\,\boldsymbol{n}_{SiO_2}$ for all $|\Re|$ peaks. This can be understood by considering the multiple-reflection model (Fig. 1a): unlike $d_2$, $d_3$ does not change with device motion. Therefore, the $d_3$ value for the highest responsivity is the one that maximizes the total reflected light intensity underneath the 2D crystal (rays 3&4), which interferes with the lights reflected from the 2D crystal surfaces (rays 1&2). This condition is met when the sacrificial

SiO$_2$ layer is completely removed ($d_3$=0), or gives completely constructive interference between rays 3&4 ($d_3=m\lambda/2\,\boldsymbol{n}_{SiO_2}$).

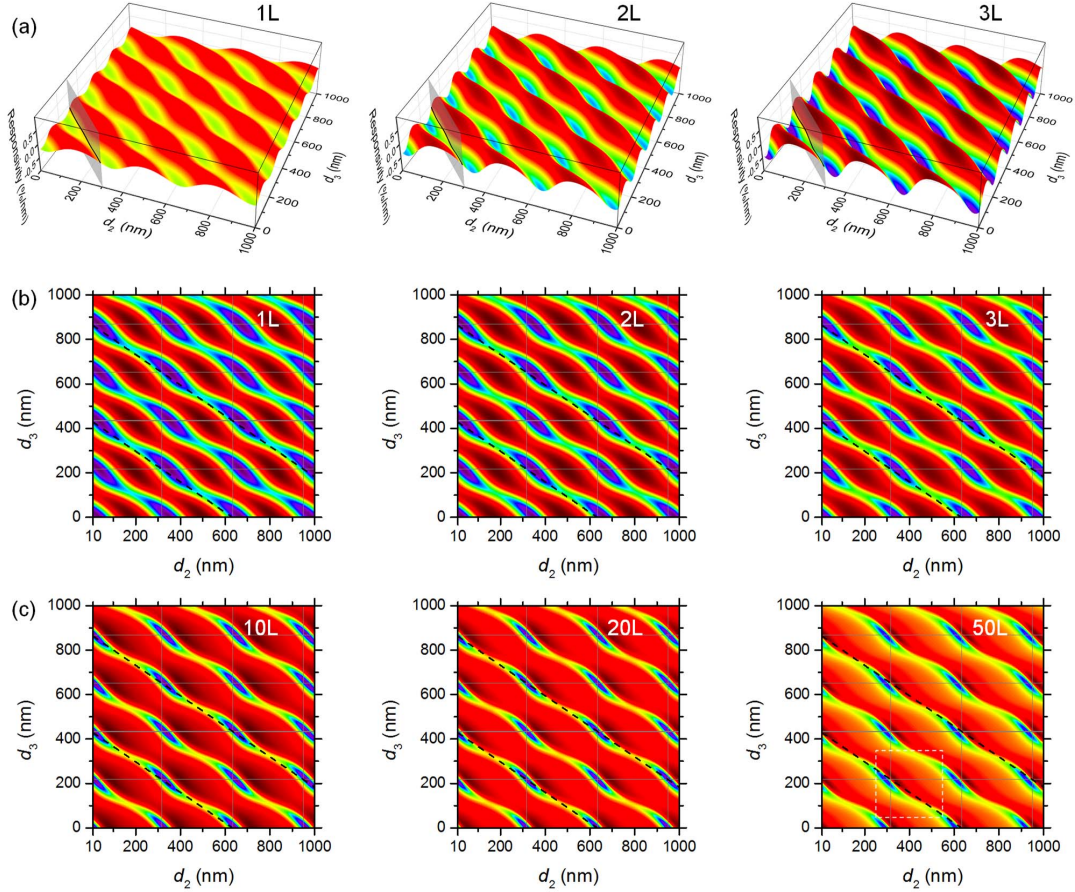

**Figure 3**. Responsivity of MoS$_2$ nanomechanical resonators as a function of vacuum gap depth ($d_2$) and oxide thickness ($d_3$). (a) 3D plots of responsivity for 1L, 2L, and 3L MoS$_2$ devices. The vertical intercepting plane represent a subset in the parameter space ($d_2+d_3$=300nm), corresponding to the data in Fig. 2. (b) 2D color plot of (a). Note that the color scale is individually optimized for each plot for easy identification of the positive (dark red) and negative (blue) responsivity peaks. (c) 2D color plot for 10, 20, and 50L devices. Locations with $d_2=n\lambda/2$ (vertical) and $d_3=m\lambda/2\,\boldsymbol{n}_{SiO_2}$ (horizontal) are shown in (b) and (c). Dashed box in (c) 50L plot represents the data range of Fig. 4. Dashed lines in (b), (c) represent positions where $d_2+d_3\boldsymbol{n}_{SiO2}$=constant.

**Dependence on Crystal Thickness.** We now focus on the dependence of $|\Re|$ peaks on $d_1$ and $d_2$, assuming $d_3$=0 and hereafter. Figure 4b shows that the $|\Re|$ peak values, both positive (wine) and negative (blue), vary non-monotonically from 1L to 50L devices: the projections onto the two vertical planes shows that $|\Re|$ (for an optimized structure) increases with thickness, reaching the highest value at 12-layer (with $d_2\approx$330nm), while further increase in thickness decreases $|\Re|$. Note that the projection on the bottom $d_1$-$d_2$ plane reproduces the front view (along $d_3$) of Fig. 4a.

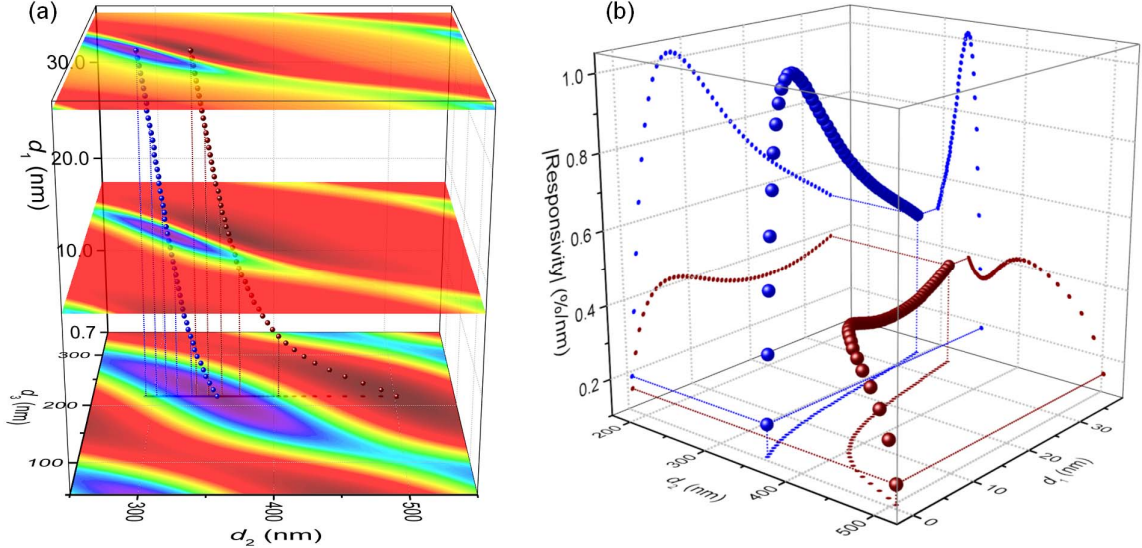

**Figure 4**. Responsivity peaks.  (a) Stacked 2D color plots (color scale individually optimized for each plot) for 1L, 25L, and 50L MoS₂ devices, with 1L to 50L responsivity peak locations (dark red: positive peaks; blue: negative peaks) extracted (also shown are their projections onto the bottom $d_2$-$d_3$ plane and selected drop lines).  The data range corresponds to the dashed box in Fig. 3c 50L plot.  (b) Magnitude of peak responsivity, both positive (wine) and negative (blue), as functions of device thickness ($d_1$) and vacuum gap depth ($d_2$), as well as data projection onto the three orthogonal planes.  Drop lines for the 1L and 50L data points are shown.

To understand this, we examine its root in reflectance $R$.  Figure 5a & b (with 3D version in 5c & d) show $R$ and $\Re$ as functions of $d_2$ for 1L to 50L ($d_1$=0.7-35nm) MoS₂ resonators (in 5c & d the thickness range is extended up to 200L).

We again use 2D color plots (Fig. 5e & f) to examine the periodic variations of $R$ and $|\Re|$.  As in Fig. 3, the period along the $d_2$ axis is strictly $\lambda/2$.  In the $d_1$ direction, $R$ oscillates between ~0-0.8, with the oscillation amplitude decrease as $d_1$ increases, approaching $R \approx 0.5$ for large $d_1$.  This period is ~60nm, or $\lambda/(2\boldsymbol{n}_{\text{MoS}_2})$, as expected from thin film optics.  The magnitude of $R$ variation decays with $d_1$ due to increased absorption for thicker crystals.  As $d_1$ becomes very large, the MoS₂ crystal becomes semi-infinite, and the optical process reduces to a single reflection at the top vacuum-MoS₂ interface (ray 1 in Fig. 1a), with $R$ approaches a constant

$$R = \left| \boldsymbol{r}_1 \right|^2 = \left| \frac{n_{vacuum} - \boldsymbol{n}_{MoS_2}}{n_{vacuum} + \boldsymbol{n}_{MoS_2}} \right|^2 \approx 0.5 \text{ and } |\Re| \text{ approaches 0 (Fig. 5c\&d).}$$

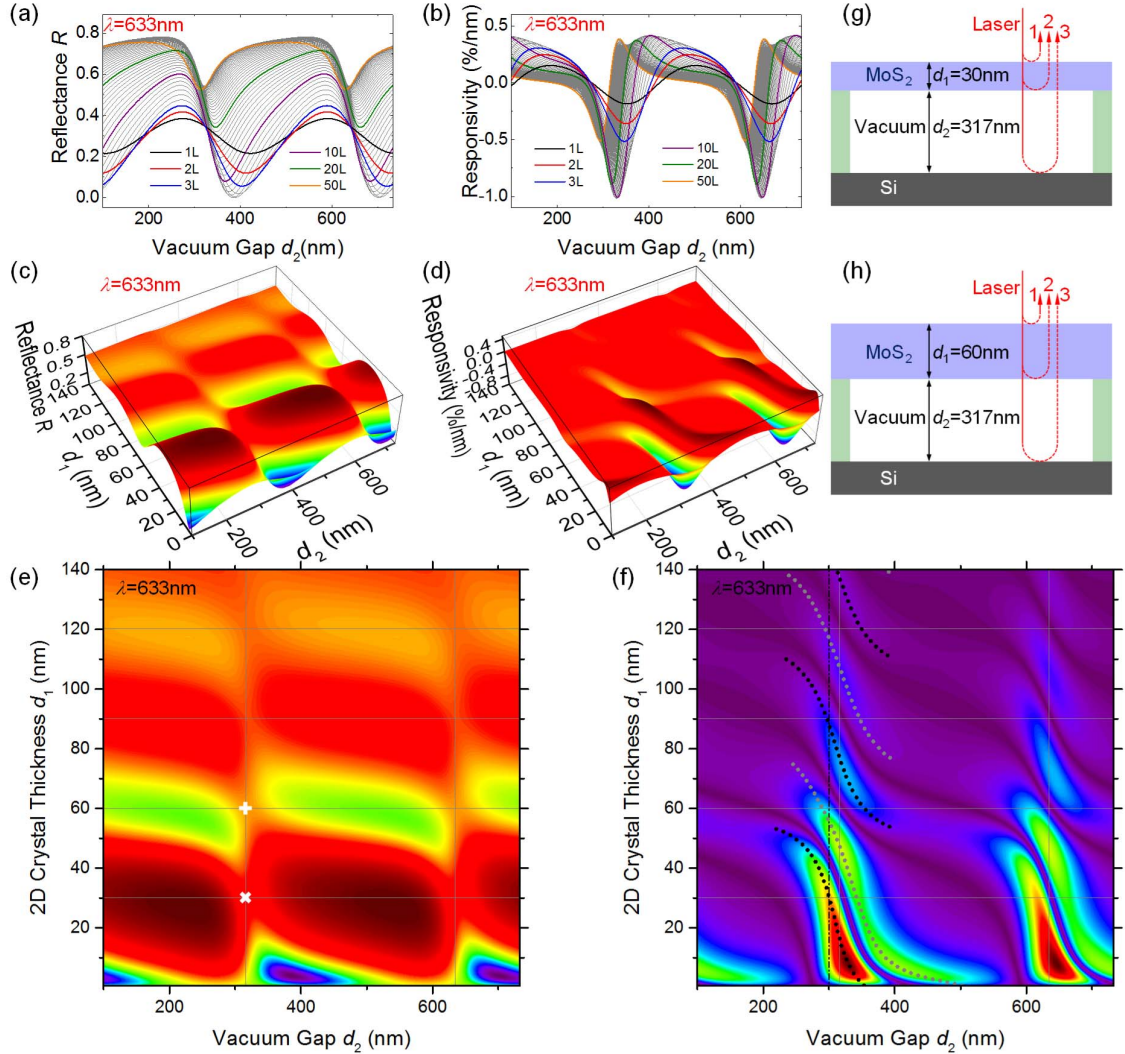

**Figure 5**. Origins of the responsivity variation. (a) Reflectance and (b) responsivity as functions of vacuum gap depth $d_2$ for 1L to 50L MoS$_2$ devices. Curves for selected thicknesses are color highlighted (indicated in legends). (c) 3D plots of reflectance $R$ and (d) responsivity $\Re$ as functions of device thickness ($d_1$) and vacuum gap depth ($d_2$) for 1L to 200L MoS$_2$ devices. (e) and (f) are 2D color plots of (c) and (d), respectively, except that $|\Re|$ is plotted instead of $\Re$, thus both positive and negative peaks both appear in the same direction on the color scale (dark purple: no responsivity; dark red: high $|\Re|$). Dotted curves represent the positive (grey) and negative (black) responsivity peaks. Locations with $d_1$=m$\lambda$/(4$\boldsymbol{n}_{MoS2}$) and $d_2$=n$\lambda$/2 are shown as gridlines. A common device geometry ($d_2$=300nm) is indicated by the vertical dash-dot line. (g) Example interferometry condition when $d_1$=(1/2+m)$\lambda$/(2$\boldsymbol{n}_{MoS2}$) and $d_2$=n$\lambda$/2 ("x" symbol in (e)). (h) Example interferometry condition when $d_1$=m$\lambda$/(2$\boldsymbol{n}_{MoS2}$) and $d_2$=n$\lambda$/2 ("+" symbol in (e)).

Close examination shows that $R$ values peaks at $d_1$=(1/2+$m$)$\lambda$/(2$\boldsymbol{n}_{MoS2}$), *e.g.*, 30nm, 90nm…($m$=1, 2…), and takes minimum at $d_1$=$m\lambda$/(2$\boldsymbol{n}_{MoS2}$), *e.g.*, 60nm (horizontal lines in Fig. 5e). Interestingly, intercepts with $d_2$=$m\lambda$/2 (vertical lines in Fig. 5e) represent local $R$ minima along $d_1=(1/2+m)\lambda/(2\boldsymbol{n}_{\text{MoS2}})$, and local maxima along $d_1=m\lambda/(2\boldsymbol{n}_{\text{MoS2}})$. This can be understood through multi-ray interference (Fig. 5g & h): to the first order we consider the interference between light paths that only experience one reflection: path 1-3, with light intensities $I_1>I_2>I_3$. The relative phases between the paths include contributions from both reflection ($\pi$ for path 1 & 3, 0 for path 2) and path length. When $d_1=(1/2+m)\lambda/(2\boldsymbol{n}_{\text{MoS2}})$ (Fig. 5g), the path length induced phase difference between paths 2 and 1 is $\pi$, resulting in a total relative phase of 0, thus overall maximizing $R$. At this $d_1$, when $d_2=m\lambda/2$, path 3 is out of phase with path 1 and 2, resulting in the local minima of $R$. Similarly (Fig. 5h), When $d_1=m\lambda/(2\boldsymbol{n}_{\text{MoS2}})$, path 2 is out of phase with path 1 (thus minimizing $R$), while $d_2=m\lambda/2$ now makes path 3 in phase with path 1, resulting in the local maxima of $R$.

**Optimization Pathways.** The $|\Re|$ map (Fig. 5f) provides a design guideline for choosing the optimal device structure (in terms of interferometric detection of device motion) for any given MoS$_2$ crystal thickness. The plots show that for $\lambda=633$nm, the 300nm-SiO$_2$-on-Si substrate widely used in 2D crystal research is a good choice: it provides good optical contrast for identifying thin crystals[36], and once etched out, the resulting ~300nm vacuum cavity (dotted line in Fig. 5f) provides good responsivity for a wide range of MoS$_2$ thicknesses. One can further tune the "effective" cavity depth $d_2$ by not fully removing SiO$_2$. As shown in Fig. 3b, $\Re$ varies slowly along $d_2+d_3\boldsymbol{n}_{\text{SiO2}}$=constant (dashed lines in Fig. 3b,c). So for example, in a recent work[10], when 250nm of 290nm SiO$_2$ is etched, the "effective" cavity depth is 250+40×1.457=308.3nm, close to the optimal $d_2$ value, allowing the observation of thermomechanical motion in MoS$_2$ resonators with different thicknesses.

**Dependence on Wavelength.** We now examine the effect from laser wavelength. We choose a few representative wavelengths: 633nm (red), 532nm (green), and 405nm (blue), all among the most commonly used lasers in labs, and cover the visible range.

Figure 6 a-c shows the results for $\lambda=532$nm. The overall pattern is similar to $\lambda=633$nm, with the main difference in the spatial periodicity in both $d_2$ and $d_1$ directions (as $\lambda$ changes). One consequence is that the magnitude of $\partial R/\partial d_2$ (*i.e.*, $|\Re|$) increases as the $d_2$ axis effectively rescales (*e.g.*, negative $\Re$ peaks in Fig. 6a bottom pane compared with those in Fig. 5b).

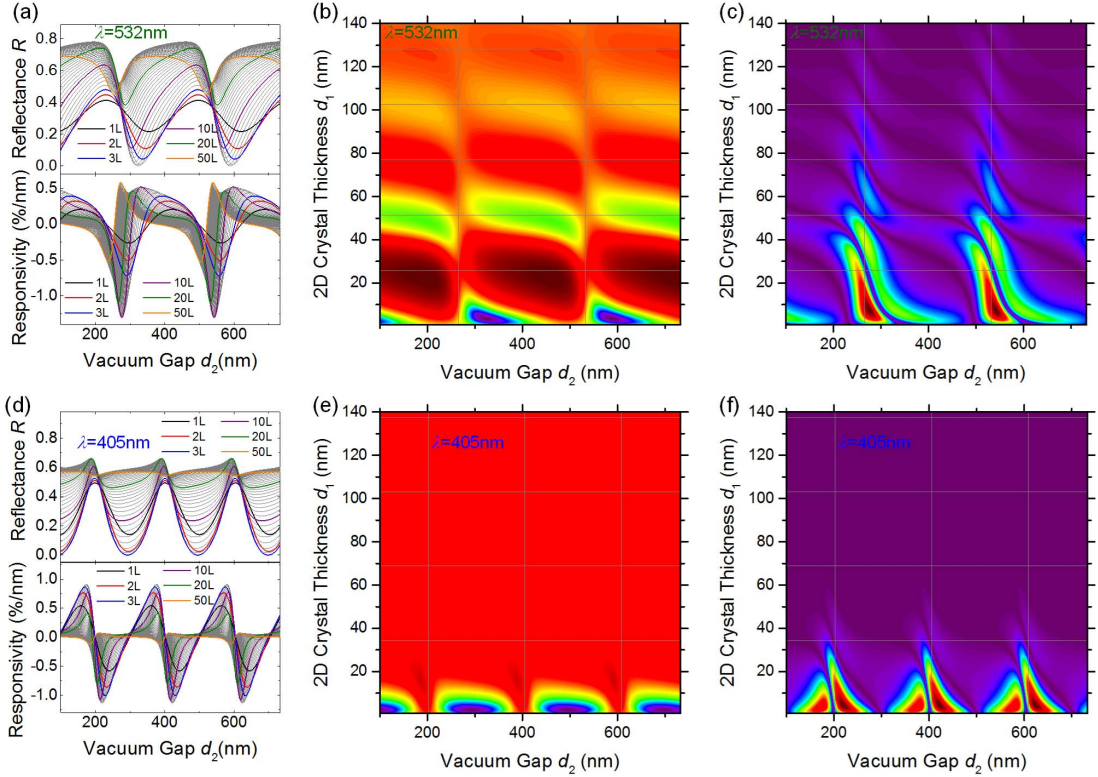

**Figure 6**. Interferometry at 532nm (a-c) and 405nm (d-f) wavelengths. (a,d) Reflectance (top) and responsivity (bottom) as functions of vacuum gap depth ($d_2$) for 1L to 50L MoS$_2$ devices. Curves for selected thicknesses are color highlighted (as indicated in legends). (b,e) 2D color plots of reflectance $R$ and (c,f) magnitude of responsivity $|\Re|$ as functions of device thickness ($d_1$) and vacuum gap depth ($d_2$) for 1L to 200L MoS$_2$ devices (in $R$ maps: dark purple: low $R$; dark red: high $R$; in $|\Re|$ maps: dark purple: no responsivity; dark red: high $|\Re|$). Locations with $d_1 = m\lambda/(4n_{MoS2})$ and $d_2 = n\lambda/2$ are shown as gridlines.

When $\lambda$ further reduces to 405nm (Fig. 6 d-f), additional effect rises: the optical absorption becomes much stronger such that the decay in $d_1$ direction is significant, and beyond ~50-layer there is little responsivity (as little light can penetrate the 2D crystal).

**Different 2D Crystals.** Figure 7 shows the results for 1-layer to 200-layer graphene and h-BN under 633nm illumination. While the periodicity in the vacuum gap depth ($d_2$) direction is preserved, one clear contrast with MoS$_2$ is the "disappearance" of periodicity along $d_1$. This directly results from the materials' low indices of refraction (compared to MoS$_2$, see Methods), making $\lambda/(2n_{2D}) > 200$-layer thickness, and thus periodicity along $d_1$ is not visible in the plots. There is also an important contrast between the graphene and h-BN cases: for graphene, the $R$ modulation (with $d_2$) is most pronounced for devices below 50nm (~150L), which consequently exhibit higher $|\Re|$; while for h-BN, the $R$ modulation (and thus $|\Re|$) monotonically increases with $d_1$ within the entire plot range. This manifests the effect from band structure. For graphene (0eV bandgap), optical absorption increases quickly with thickness, thus thinner devices (< 150L) exhibit higher responsivity. In contrast, h-BN has large bandgap (~5eV) and thus minimal absorption at 633nm; together with the relatively low refractive index (thus low reflectivity)[40], there is little *R* modulation as h-BN device vibrates, unless the crystal is sufficiently thick to induce sizable absorption. The results show that for mono-and few-layer h-BN, (even with optimized device geometry) $|\Re|$ is orders of magnitude lower than for graphene and MoS$_2$, and only multilayer (>20L, green curve in Fig. 7d) h-BN has comparable $|\Re|$ values as monolayer MoS$_2$ or graphene.

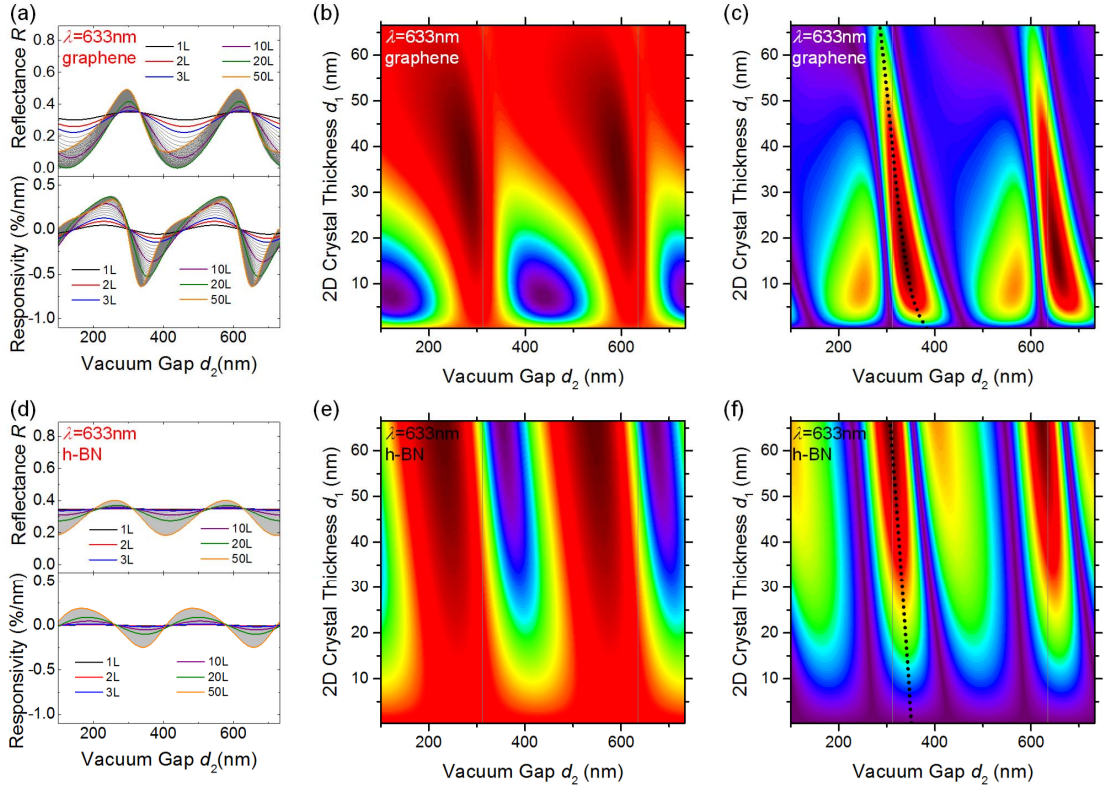

**Figure 7**. Reflectance and responsivity of graphene (a-c) and h-BN (d-f) devices with $\lambda$=633nm. (a,d) Reflectance (top) and responsivity (bottom) as functions of vacuum gap depth ($d_2$) for 1L to 50L graphene/h-BN devices. Curves for selected thicknesses are color highlighted (as indicated in legends). The vertical scales are the same as in Fig. 5a&b for easy comparison. (b,e) 2D color plots of reflectance *R* and (c,f) magnitude of responsivity $|\Re|$ as functions of device thickness ($d_1$) and vacuum gap depth ($d_2$) for 1L to 200L graphene and h-BN devices (in *R* maps: dark purple: low *R*; dark red: high *R*; in $|\Re|$ maps: dark purple: no responsivity; dark red: high $|\Re|$). Dotted curves represent the responsivity peaks. Locations with $d_2=n\lambda/2$ are shown as gridlines.

**Quantitative Design Guideline for Optimizing Responsivity.** We finally summarize the optimized responsivity and corresponding device structures for the different 2D materials for the three widely employed wavelengths. Figure 8 (a) shows the highest achievable $|\Re|$ values for 1L to 200L graphene, h-BN, and MoS$_2$ devices under 405nm, 532nm, and 633nm illumination. The line style represents different 2D material and line color corresponds to each wavelength. The results show that towards the monolayer limit, $MoS_2$ devices can have the highest $|\Re|$ among these 2D materials for the three wavelengths (1L to 14L for 405nm; 1L to 33L for 532nm; 1L to 37L for 633nm). As the number of layers increases, the highest $|\Re|$ values are found in graphene structures (405nm: 15L to 100L; 532nm: 34L to 116L; 633nm: 38L to 112L). In even thicker structures, up to 200L, h-BN devices offer the highest $|\Re|$ (405nm: $\geq$101L; 532nm: $\geq$117L; 633nm: $\geq$113L). The color bars on the top of Fig. 8(a) summarizes the thickness range in which each particular 2D material exhibits the highest $|\Re|$ in their respective optimized device geometry. Figure 8(b) shows the optimal vacuum gap depth $d_2$ for achieving the highest $|\Re|$ values discussed above, providing a clear design guideline for NEMS devices based on these 2D materials.

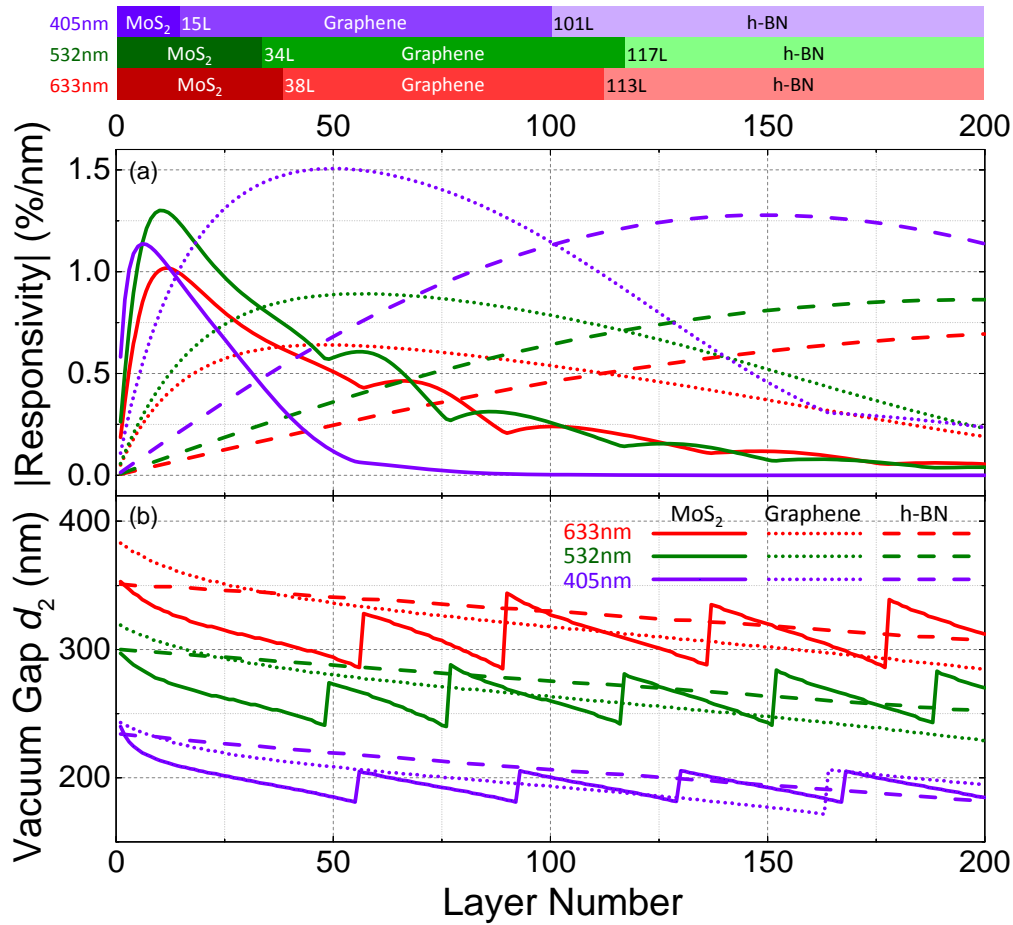

**Figure 8**. Summary of 1-layer to 200-layer structures with (a) optimized responsivity values and (b) their corresponding structures for $MoS_2$ (solid lines), graphene (dotted lines), and h-BN (dashed lines) devices with $\lambda$=405nm (purple lines), 532nm (green lines), and 633nm (red lines). The optimal $d_2$ values around $\lambda/2$ are shown (the results are periodic about $m\lambda/2$). The saw teeth patterns in $d_2$ curves (mostly visible for $MoS_2$) result from the $|\Re|$ peaks alternating between positive and negative peaks, which can also be noticed in the $|\Re|$ plot in (a). The color bars on top illustrates the ranges of thickness (in number of layers) for each 2D material to have the highest $|\Re|$ with optimized device structure, with number of layers labeled for each new thickness range.

The physical origin of the findings (that as thickness varies, different 2D crystals attain their highest motion transduction responsivity conditions with different device configurations) can be intuitively understood by considering two mechanisms of light-device interaction: reflection and absorption. Multireflection (Fig. 1a) causes interference between different optical paths, which generates the detailed interferometric effects and signals and thus determines the motion transduction responsivity. Inside the multilayer structure (see Fig. 1a), in the regime that the 2D material layer (thickness $d_1$) only reflects less than a few percent of incident light (as often found in ultrathin samples), if a 2D flake gives comparatively more reflection, it leads to stronger interference effects inside the vacuum gap (depth $d_2$) and thus higher motion transduction responsivity (as $d_2$ varies). Further, such multireflection of light within the layered structure (Fig. 1a) creates a spatially-varying optical field in the vacuum gap underneath the 2D crystal. As the vacuum gap depth varies, the light intensity at the 2D flake changes, and the finite absorption in the 2D material causes change in the total light intensity. In the regime that the absorption (here we specifically refer to the percentage of light intensity absorbed when passing through a 2D flake) of the 2D flake is only a few percent or less (as often found in ultrathin samples) such that sufficient light enters the vacuum gap (to form a spatially-varying optical field), 2D flakes with greater absorption can engender greater responsivity.

Larger index of refraction $n$ leads to greater reflection. For all the wavelengths in this study, $n_{MoS2} > n_{graphene} > n_{h-BN}$. In the visible spectrum, $MoS_2$ has greater absorption than graphene (see Methods), while h-BN absorbs the least as its bandgap corresponds to ultraviolet. Therefore, mono- and few-layer $MoS_2$ (h-BN) devices have the greatest (smallest) responsivity as the flake is most (least) reflecting and absorbing (as shown in the left end of Fig. 8a).

As the thickness (and thus total absorption) increases beyond just a few percent (as often the case for many-layer or thin film samples), the dependence of responsivity on 2D flake reflectance and absorption can change. High reflection causes less light entering the vacuum gap (for interferometry), and stronger absorption leads to less total reflected light (thus less intensity available for modulation by device motion); both reduce the responsivity for multilayer 2D flakes. Therefore $MoS_2$ (h-BN), the most (least) reflecting and absorbing among the three crystals, is the first (last) to experience such effect: beyond a given thickness, responsivity starts to decrease as number of layers further increases, as seen in the solid (dashed) curves in Fig 8a.

In conclusion, we have systemically investigated interferometric motion detection in 2D nanomechanical devices based on atomic layer crystals. We have quantitatively elucidated and graphically visualized the dependences of motion responsivity upon parameters in device structure, probing wavelength, and type of 2D material. We find that the highest responsivity may be achieved with no oxide layer at the bottom of the vacuum gap, and the optimal vacuum gap varies (with crystal thickness) around $m\lambda/2$; specifically, when using 633nm He-Ne laser, the ~300nm-$SiO_2$-on-Si substrates commonly used in 2D research (and the resulting vacuum gap) offer close-to-optimal motion responsivity for several 2D crystals over a wide range of thickness. We also illustrate the trade-off between enhancing responsivity and increasing absorption when using short wavelengths, and show that different types of 2D layered materials exhibit different patterns in the same parameter space due to their different band structure and dielectric constants. The optimization pathways shown in our results provide a complete design guideline for building 2D nanomechanical devices with the highest achievable optical transduction efficiency, which can significantly improve the signal-to-noise ratio and enhance device performance such as dynamic range and sensitivity. This can in turn help enable new functions and high performance in emerging applications, *e.g.*, future fiber-optic, near-field, and on-chip interferometric schemes with ultra-sensitive signal detection and processing using 2D nanodevices.

**Table 1**. Optimal Geometry for 1L, 2L and 3L Graphene, h-BN, and MoS$_2$ Devices.

| *Thickness* | **1L** | | **2L** | | **3L** | | **# of Layers Free to Vary for Highest Possible Responsivity** | | | |
|---|---|---|---|---|---|---|---|---|---|---|
| *Material* | Vacuum Gap $d_2$ (nm) | Optimal $|\Re|$ (%/nm) | Vacuum Gap $d_2$ (nm) | Optimal $|\Re|$ (%/nm) | Vacuum Gap $d_2$ (nm) | Optimal $|\Re|$ (%/nm) | Number of Layers | Crystal Thickness $d_1$ (nm) | Vacuum Gap $d_2$ (nm) | Optimal $|\Re|$ (%/nm) |
| *Graphene* | 383 | 0.0508 | 381 | 0.0973 | 378 | 0.140 | 48 | 16.08 | 337 | 0.640 |
| *h-BN* | 351 | 0.00503 | 351 | 0.0101 | 350 | 0.0151 | 240 | 79.92 | 297 | 0.714 |
| *MoS$_2$* | 353 | 0.186 | 350 | 0.360 | 347 | 0.515 | 12 | 8.4 | 329 | 1.016 |

## Methods

The indices of refraction (complex values) are obtained from a number of references. In some cases, instead of the complex refractive index $\boldsymbol{n}$=$n$-$i\kappa$ ($\kappa$ is also called the "extinction coefficient"), the values are given in the form of complex dielectric constant, *i.e.* relative permittivity: $\boldsymbol{\varepsilon}$=$\varepsilon_1$+$i\varepsilon_2$, which is related to the index of refraction through $\boldsymbol{\varepsilon} = \boldsymbol{n}^2 = \left(n - i\kappa\right)^2$. One can calculate $n$ and $\kappa$ using $n = \sqrt{\dfrac{\sqrt{\varepsilon_1^2 + \varepsilon_2^2} + \varepsilon_1}{2}}$ and $\kappa = \sqrt{\dfrac{\sqrt{\varepsilon_1^2 + \varepsilon_2^2} - \varepsilon_1}{2}}$. For layered materials, we verify that the values are for incident light normal to the basal plane (*i.e.*, parallel to the crystalline c axis). Below we list the values used in this study:

Vacuum: $n_{vacuum}$=1;

Silicon[39]: $\boldsymbol{n}_{Si}$=3.882-0.019$i$ (633nm); 4.15-.044$i$ (532nm); 5.420-0.329$i$ (405nm);

Silicon dioxide[39,41] (note that the value is a real number at this wavelength[33]): $n_{SiO2}$=1.457 (633nm);

Molybdenum disulfide[10,42]: $\boldsymbol{n}_{MoS2}$=5.263-1.14$i$ (633nm); 5.185-1.121$i$ (532nm); 3.868-3.231$i$ (405nm); monolayer thickness 0.7nm;

Graphene[39]: $\boldsymbol{n}_{graphene}$=2.86-1.73$i$ (633nm); 2.67-1.34$i$ (532nm); 2.62-1.29$i$ (405nm); monolayer thickness 0.335nm;

Hexagonal boron nitride[39, 43]: $\boldsymbol{n}_{\text{h-BN}}$=1.605-0.19$i$ (633nm); 1.585-0.22$i$ (532nm); 1.55-0.25$i$ (405nm); monolayer thickness 0.333nm.

We note that the refractive index could be layer-dependent for certain materials, *e.g.*, the value for monolayer could be different from that of the bulk[44]. Earlier study suggests that such difference is expected to be insignificant, as the optical response of 2D layered material with light incident normal to the basal plane is dominated by in-plane electromagnetic response, which is similar in few-layer structures and in bulk[33]. We therefore use the same value for all the device thicknesses in calculation, as in previous work[33,35,36].

We also note that different sources in the literature may give different refractive index values[39,43,45]. While such quantitative differences can lead to changes in the numerical values (and causes the patterns in the figures to slightly shift), all the results remain qualitatively unchanged, and all the physical arguments and interpretations remain valid.

While monolayer graphene has zero bandgap and exhibit strong absorption (percentage of light intensity absorbed when passing through a 2D flake, also sometimes called "absorbance" in 2D literature) over a wide spectrum range[46,47], monolayer $MoS_2$ has greater absorption in the visible range due to interband transitions and higher density of states[48]. This is also manifested in their complex indices of refraction: the absorption coefficient $\alpha=4\pi\kappa/\lambda$ is proportional to the extinction coefficient $\kappa$, and in the ultrathin limit the exponential dependence of absorption on flake thickness $d_1$ reduces to a linear relation: absorption $\propto \alpha d_1 \propto \kappa d_1$. Using monolayer thickness and $\kappa$ values of $MoS_2$ and graphene, we estimate that monolayer absorption of $MoS_2$ is ~40% greater than that of graphene at 633nm, consistent with measured values[46,48].

Table 1 summarizes the optimized device geometry for 1L, 2L, and 3L nanodevices based on 2D materials under 633nm illumination. See Supplementary Information for complete 1L through 200L data.

# REFERENCES

1 Geim, A. K. & Novoselov, K. S. The rise of graphene. *Nat. Mater.* **6,** 183–191 (2007).

2 Geim, A. K. Graphene: Status and prospects. *Science* **324,** 1530–1534 (2009).

3 Butler, S. Z. *et al.* Progress, challenges, and opportunities in two-dimensional materials beyond graphene. *ACS Nano* **7,** 2898–2926 (2013).

4 Wang, Q. H., Kalantar-Zadeh, K., Kis, A., Coleman, J. N. & Strano, M. S. Electronics and optoelectronics of two-dimensional transition metal dichalcogenides. *Nat. Nanotechnol.* **7,** 699–712 (2012).

5 Fiori, G. *et al.* Electronics based on two-dimensional materials. *Nat. Nanotechnol.* **9,** 768–779 (2014).

6 Lee, C., Wei, X., Kysar, J. W. & Hone, J. Measurement of the elastic properties and intrinsic strength of monolayer graphene. *Science.* **321,** 385–388 (2008).

7 Feng, P. X.-L. *et al.* Two-dimensional nanoelectromechanical systems (2D NEMS) via atomically-thin semiconducting crystals vibrating at radio frequencies, in Tech. Dig. Int. Electron Dev. Meet. (IEDM'2014), Paper No. 8.1, pp. 196-199, San Francisco, CA, December 15-17 (2014).

8 Barton, R. A., Parpia, J. & Craighead, H. G. Fabrication and performance of graphene nanoelectromechanical systems. *J. Vac. Sci. Technol. B* **29,** 050801 (2011).

9 Sengupta, S., Solanki, H. S., Singh, V., Dhara, S. & Deshmukh, M. M. Electromechanical resonators as probes of the charge density wave transition at the nanoscale in $NbSe_2$. *Phys. Rev. B* **82,** 155432 (2010).

10 Lee, J., Wang, Z., He, K., Shan, J. & Feng, P. X.-L. High frequency $MoS_2$ nanomechanical resonators. *ACS Nano* **7,** 6086–6091 (2013).

11 Wang, Z. *et al.* Black phosphorus nanoelectromechanical resonators vibrating at very high frequencies. *Nanoscale* **7,** 877-884 (2015).

12 Chen, C. *et al.* Performance of monolayer graphene nanomechanical resonators with electrical readout. *Nat. Nanotechnol.* **4,** 861–867 (2009).

13 van der Zande, A. M. *et al.* Large-scale arrays of single-layer graphene resonators. *Nano Lett.* **10,** 4869–4873 (2010).

14 Xu, Y. *et al.* Radio frequency electrical transduction of graphene mechanical resonators. *Appl. Phys. Lett.* **97,** 243111 (2010).

15 Lee, S. *et al.* Electrically integrated SU-8 clamped graphene drum resonators for strain engineering. *Appl. Phys. Lett.* **102,** 153101 (2013).

16 Garcia-Sanchez, D. *et al.* Imaging mechanical vibrations in suspended graphene sheets. *Nano Lett.* **8,** 1399–1403 (2008).

17 Bunch, J. S. *et al.* Electromechanical Resonators from Graphene Sheets. *Science* **315,** 490–493 (2007).

18 Barton, R. A. *et al.* High, size-dependent quality factor in an array of graphene mechanical resonators. *Nano Lett.* **11,** 1232–1236 (2011).

19 Barton, R. A. *et al.* Photothermal self-oscillation and laser cooling of graphene optomechanical systems. *Nano Lett.* **12,** 4681–4686 (2012).

20 Michelson, A. A. & Morley, E. W. On the relative motion of the Earth and the luminiferous ether. *Am. J. Sci.* **s3-34,** 333–345 (1887).

21 Abbott, B. P. *et al.* (LIGO Scientific Collaboration and Virgo Collaboration) Observation of gravitational waves from a binary black hole merger. *Phys. Rev. Lett.* **116,** 061102 (2016).

22 Rugar, D., Budakian, R., Mamin, H. J. & Chui, B. W. Single spin detection by magnetic resonance force microscopy. *Nature* **430,** 329–332 (2004).

23 Aasi, J. *et al.* Enhanced sensitivity of the LIGO gravitational wave detector by using squeezed states of light. *Nat. Photonics* **7,** 613–619 (2013).

24 Abadie, J. *et al.* (LIGO Scientific Collaboration) A gravitational wave observatory operating beyond the quantum shot-noise limit. *Nat. Phys.* **7,** 962–965 (2011).

25 Ilic, B., Krylov, S., Aubin, K., Reichenbach, R. & Craighead, H. G. Optical excitation of nanoelectromechanical oscillators. *Appl. Phys. Lett.* **86,** 193114 (2005).

26 Hiebert, W. K., Vick, D., Sauer, V. & Freeman, M. R. Optical interferometric displacement calibration and thermomechanical noise detection in bulk focused ion beam-fabricated nanoelectromechanical systems. *J. Micromech. Microeng.* **20,** 115038 (2010).

27 Karabalin, R. B. *et al.* Piezoelectric nanoelectromechanical resonators based on aluminum nitride thin films. *Appl. Phys. Lett.* **95,** 103111 (2009).

28 Kouh, T., Karabacak, D., Kim, D. H. & Ekinci, K. L. Diffraction effects in optical interferometric displacement detection in nanoelectromechanical systems. *Appl. Phys. Lett.* **86,** 013106 (2004).

29 Nichol, J. M., Hemesath, E. R., Lauhon, L. J. & Budakian, R. Displacement detection of silicon nanowires by polarization-enhanced fiber-optic interferometry. *Appl. Phys. Lett.* **93,** 193110 (2008).

30 Belov, M. *et al.* Mechanical resonance of clamped silicon nanowires measured by optical interferometry. *J. Appl. Phys.* **103,** 074304 (2008).

31 Karabacak, D., Kouh, T. & Ekinci, K. L. Analysis of optical interferometric displacement detection in nanoelectromechanical systems. *J. Appl. Phys.* **98,** 124309 (2005).

32 Wang, Z., Lee, J. & Feng, P. X.-L. Spatial mapping of multimode Brownian motions in high-frequency silicon carbide microdisk resonators. *Nat. Commun.* **5,** 5158 (2014).

33 Blake, P. *et al.* Making graphene visible. *Appl. Phys. Lett.* **91,** 063124 (2007).

34 Gorbachev, R. V. *et al.* Hunting for monolayer boron nitride: optical and raman signatures. *Small* **7,** 465–468 (2011).

35 Castellanos-Gomez, A., Agraït, N. & Rubio-Bollinger, G. Optical identification of atomically thin dichalcogenide crystals. *Appl. Phys. Lett.* **96,** 213116 (2010).

36 Benameur, M. M. *et al.* Visibility of dichalcogenide nanolayers. *Nanotechnology* **22,** 125706 (2011).

37 Anders, H. *Thin Films in Optics* (Focal Press, 1967).

38 Yang, R., Zheng, X., Wang, Z., Miller, C. J. & Feng, P. X.-L. Multilayer $MoS_2$ transistors enabled by a facile dry-transfer technique and thermal annealing. *J. Vac. Sci. Technol. B* **32,** 061203 (2014).

39 E. D. Palik, *Handbook of Optical Constants of Solids* (Academic Press, 1998).

40 Golberg, D. *et al.* Boron nitride nanotubes and nanosheets. *ACS Nano* **4,** 2979–2993 (2010).

41 Malitson, I. H. Interspecimen comparison of the refractive index of fused silica. *J. Opt. Soc. Am.* **55,** 1205 (1965).

42 Beal, A. R. & Hughes, H. P. Kramers-Kronig analysis of the reflectivity spectra of 2H-$MoS_2$, 2H-$MoSe_2$ and 2H-$MoTe_2$. *J. Phys. C Solid State Phys.* **12,** 881 (1979).

43 Hoffman, D. M., Doll, G. L. & Eklund, P. C. Optical properties of pyrolytic boron nitride in the energy range 0.05-10 eV. *Phys. Rev. B* **30,** 6051–6056 (1984).

44 Zhang, H. *et al.* Measuring the refractive index of highly crystalline monolayer $MoS_2$ with high confidence. *Sci. Rep.* **5,** 8440 (2015).

45 Schubert, M. *et al.* Anisotropy of boron nitride thin-film reflectivity spectra by generalized ellipsometry. *Appl. Phys. Lett.* **70,** 1819–1821 (1997).

46 Mak, K. F. *et al.* Measurement of the optical conductivity of graphene, *Phys. Rev. Lett.* **101,** 196405 (2008).

47 Kuzmenko, A. B., van Heumen, E., Carbone, F. & van der Marel, D., Universal optical conductance of graphite, *Phys. Rev. Lett.* **100,** 117401 (2008).

48 Mak, K. F. *et al.* Tightly bound trions in monolayer $MoS_2$. *Nat. Mat.* **12,** 207–211 (2013).

**Acknowledgements**
We thank the support from Case School of Engineering, National Academy of Engineering (NAE) Grainger Foundation Frontier of Engineering (FOE) Award (FOE2013-005), and National Science Foundation CAREER Award (ECCS #1454570). We thank Jaesung Lee and Keliang He for helpful discussions.

**Author contributions**
Z.W. performed calculations and analyzed the data. Z.W. and P.X.-L.F. wrote the manuscript. Both authors discussed about the results and iterated on revising the manuscript. P.X.-L.F. conceived and supervised the study.

**Additional information**
**Supplementary information** accompanies this paper at http://www.nature.com/scientificreports
**Competing financial interest:** the authors declare no competing financial interests.

# Interferometric Motion Detection in Atomic Layer 2D Nanostructures:
# Visualizing Signal Transduction Efficiency and Optimization Pathways

Zenghui Wang, Philip X.-L. Feng[*]

*Department of Electrical Engineering & Computer Science,*

*Case School of Engineering, Case Western Reserve University,*

*10900 Euclid Avenue, Cleveland, OH 44106, USA*

**Table S1**. Optimal Geometry for 1L– 200L Graphene, h-BN, and $MoS_2$ Devices.

| 2D Material | Graphene | | | h-BN | | | $MoS_2$ | | |
|---|---|---|---|---|---|---|---|---|---|
| Number of Layers | Crystal Thickness $d_1$ (nm) | Vacuum Gap $d_2$ (nm) | Optimal $|\Re|$ (%/nm) | Crystal Thickness $d_1$ (nm) | Vacuum Gap $d_2$ (nm) | Optimal $|\Re|$ (%/nm) | Crystal Thickness $d_1$ (nm) | Vacuum Gap $d_2$ (nm) | Optimal $|\Re|$ (%/nm) |
| 1 | 0.335 | 383 | 0.051 | 0.333 | 351 | 0.005 | 0.7 | 353 | 0.186 |
| 2 | 0.67 | 381 | 0.097 | 0.666 | 351 | 0.010 | 1.4 | 350 | 0.360 |
| 3 | 1.005 | 378 | 0.140 | 0.999 | 350 | 0.015 | 2.1 | 347 | 0.515 |
| 4 | 1.34 | 376 | 0.179 | 1.332 | 350 | 0.020 | 2.8 | 344 | 0.648 |
| 5 | 1.675 | 375 | 0.216 | 1.665 | 350 | 0.025 | 3.5 | 342 | 0.757 |
| 6 | 2.01 | 373 | 0.249 | 1.998 | 350 | 0.030 | 4.2 | 339 | 0.844 |
| 7 | 2.345 | 371 | 0.280 | 2.331 | 350 | 0.035 | 4.9 | 337 | 0.909 |
| 8 | 2.68 | 369 | 0.309 | 2.664 | 349 | 0.040 | 5.6 | 335 | 0.956 |
| 9 | 3.015 | 368 | 0.335 | 2.997 | 349 | 0.045 | 6.3 | 333 | 0.988 |
| 10 | 3.35 | 366 | 0.360 | 3.33 | 349 | 0.050 | 7 | 332 | 1.007 |
| 11 | 3.685 | 365 | 0.383 | 3.663 | 349 | 0.055 | 7.7 | 330 | 1.016 |
| 12 | 4.02 | 364 | 0.404 | 3.996 | 349 | 0.060 | 8.4 | 329 | 1.016 |
| 13 | 4.355 | 363 | 0.424 | 4.329 | 349 | 0.065 | 9.1 | 327 | 1.011 |
| 14 | 4.69 | 361 | 0.442 | 4.662 | 348 | 0.070 | 9.8 | 326 | 1.000 |
| 15 | 5.025 | 360 | 0.459 | 4.995 | 348 | 0.075 | 10.5 | 325 | 0.987 |
| 16 | 5.36 | 359 | 0.474 | 5.328 | 348 | 0.080 | 11.2 | 324 | 0.970 |
| 17 | 5.695 | 358 | 0.489 | 5.661 | 348 | 0.085 | 11.9 | 322 | 0.952 |
| 18 | 6.03 | 357 | 0.502 | 5.994 | 348 | 0.090 | 12.6 | 321 | 0.933 |
| 19 | 6.365 | 356 | 0.515 | 6.327 | 347 | 0.095 | 13.3 | 320 | 0.914 |
| 20 | 6.7 | 355 | 0.527 | 6.66 | 347 | 0.100 | 14 | 319 | 0.894 |
| 21 | 7.035 | 354 | 0.537 | 6.993 | 347 | 0.105 | 14.7 | 318 | 0.874 |

[*]Corresponding Author. Email: philip.feng@case.edu

| 2D Material | Graphene | | | h-BN | | | MoS$_2$ | | |
|---|---|---|---|---|---|---|---|---|---|
| Number of Layers | Crystal Thickness $d_1$ (nm) | Vacuum Gap $d_2$ (nm) | Optimal $|\Re|$ (%/nm) | Crystal Thickness $d_1$ (nm) | Vacuum Gap $d_2$ (nm) | Optimal $|\Re|$ (%/nm) | Crystal Thickness $d_1$ (nm) | Vacuum Gap $d_2$ (nm) | Optimal $|\Re|$ (%/nm) |
| 22 | 7.37 | 353 | 0.547 | 7.326 | 347 | 0.110 | 15.4 | 318 | 0.854 |
| 23 | 7.705 | 353 | 0.557 | 7.659 | 347 | 0.115 | 16.1 | 317 | 0.835 |
| 24 | 8.04 | 352 | 0.565 | 7.992 | 346 | 0.120 | 16.8 | 316 | 0.817 |
| 25 | 8.375 | 351 | 0.573 | 8.325 | 346 | 0.125 | 17.5 | 315 | 0.799 |
| 26 | 8.71 | 350 | 0.580 | 8.658 | 346 | 0.130 | 18.2 | 314 | 0.782 |
| 27 | 9.045 | 349 | 0.587 | 8.991 | 346 | 0.135 | 18.9 | 313 | 0.765 |
| 28 | 9.38 | 349 | 0.593 | 9.324 | 346 | 0.140 | 19.6 | 312 | 0.749 |
| 29 | 9.715 | 348 | 0.599 | 9.657 | 345 | 0.145 | 20.3 | 312 | 0.734 |
| 30 | 10.05 | 347 | 0.604 | 9.99 | 345 | 0.150 | 21 | 311 | 0.720 |
| 31 | 10.385 | 347 | 0.608 | 10.323 | 345 | 0.155 | 21.7 | 310 | 0.706 |
| 32 | 10.72 | 346 | 0.613 | 10.656 | 345 | 0.160 | 22.4 | 309 | 0.692 |
| 33 | 11.055 | 345 | 0.617 | 10.989 | 345 | 0.165 | 23.1 | 309 | 0.680 |
| 34 | 11.39 | 345 | 0.620 | 11.322 | 344 | 0.170 | 23.8 | 308 | 0.668 |
| 35 | 11.725 | 344 | 0.623 | 11.655 | 344 | 0.175 | 24.5 | 307 | 0.656 |
| 36 | 12.06 | 344 | 0.626 | 11.988 | 344 | 0.180 | 25.2 | 306 | 0.645 |
| 37 | 12.395 | 343 | 0.629 | 12.321 | 344 | 0.184 | 25.9 | 305 | 0.634 |
| 38 | 12.73 | 342 | 0.631 | 12.654 | 344 | 0.189 | 26.6 | 305 | 0.624 |
| 39 | 13.065 | 342 | 0.633 | 12.987 | 343 | 0.194 | 27.3 | 304 | 0.614 |
| 40 | 13.4 | 341 | 0.635 | 13.32 | 343 | 0.199 | 28 | 303 | 0.604 |
| 41 | 13.735 | 341 | 0.636 | 13.653 | 343 | 0.204 | 28.7 | 302 | 0.595 |
| 42 | 14.07 | 340 | 0.637 | 13.986 | 343 | 0.209 | 29.4 | 301 | 0.585 |
| 43 | 14.405 | 340 | 0.638 | 14.319 | 342 | 0.213 | 30.1 | 300 | 0.576 |
| 44 | 14.74 | 339 | 0.639 | 14.652 | 342 | 0.218 | 30.8 | 299 | 0.567 |
| 45 | 15.075 | 339 | 0.640 | 14.985 | 342 | 0.223 | 31.5 | 299 | 0.558 |
| 46 | 15.41 | 338 | 0.640 | 15.318 | 342 | 0.228 | 32.2 | 298 | 0.548 |
| 47 | 15.745 | 338 | 0.640 | 15.651 | 342 | 0.232 | 32.9 | 297 | 0.539 |
| 48 | 16.08 | 337 | 0.640 | 15.984 | 341 | 0.237 | 33.6 | 296 | 0.529 |
| 49 | 16.415 | 337 | 0.640 | 16.317 | 341 | 0.242 | 34.3 | 295 | 0.519 |
| 50 | 16.75 | 336 | 0.640 | 16.65 | 341 | 0.247 | 35 | 294 | 0.509 |
| 51 | 17.085 | 336 | 0.640 | 16.983 | 341 | 0.251 | 35.7 | 292 | 0.498 |
| 52 | 17.42 | 335 | 0.639 | 17.316 | 341 | 0.256 | 36.4 | 291 | 0.487 |
| 53 | 17.755 | 335 | 0.639 | 17.649 | 340 | 0.261 | 37.1 | 290 | 0.476 |
| 54 | 18.09 | 335 | 0.638 | 17.982 | 340 | 0.265 | 37.8 | 289 | 0.464 |
| 55 | 18.425 | 334 | 0.637 | 18.315 | 340 | 0.270 | 38.5 | 287 | 0.451 |
| 56 | 18.76 | 334 | 0.636 | 18.648 | 340 | 0.274 | 39.2 | 286 | 0.437 |
| 57 | 19.095 | 333 | 0.635 | 18.981 | 339 | 0.279 | 39.9 | 285 | 0.429 |
| 58 | 19.43 | 333 | 0.634 | 19.314 | 339 | 0.284 | 40.6 | 283 | 0.435 |
| 59 | 19.765 | 332 | 0.633 | 19.647 | 339 | 0.288 | 41.3 | 282 | 0.440 |
| 60 | 20.1 | 332 | 0.631 | 19.98 | 339 | 0.293 | 42 | 280 | 0.446 |
| 61 | 20.435 | 332 | 0.630 | 20.313 | 339 | 0.297 | 42.7 | 278 | 0.451 |
| 62 | 20.77 | 331 | 0.628 | 20.646 | 338 | 0.302 | 43.4 | 276 | 0.455 |
| 63 | 21.105 | 331 | 0.627 | 20.979 | 338 | 0.306 | 44.1 | 274 | 0.458 |
| 64 | 21.44 | 330 | 0.625 | 21.312 | 338 | 0.311 | 44.8 | 272 | 0.461 |

| 2D Material | Graphene | | | h-BN | | | MoS₂ | | |
|---|---|---|---|---|---|---|---|---|---|
| Number of Layers | Crystal Thickness $d_1$ (nm) | Vacuum Gap $d_2$ (nm) | Optimal $\lvert\Re\rvert$ (%/nm) | Crystal Thickness $d_1$ (nm) | Vacuum Gap $d_2$ (nm) | Optimal $\lvert\Re\rvert$ (%/nm) | Crystal Thickness $d_1$ (nm) | Vacuum Gap $d_2$ (nm) | Optimal $\lvert\Re\rvert$ (%/nm) |
| 65 | 21.775 | 330 | 0.624 | 21.645 | 338 | 0.315 | 45.5 | 270 | 0.463 |
| 66 | 22.11 | 330 | 0.622 | 21.978 | 337 | 0.320 | 46.2 | 267 | 0.464 |
| 67 | 22.445 | 329 | 0.620 | 22.311 | 337 | 0.324 | 46.9 | 264 | 0.464 |
| 68 | 22.78 | 329 | 0.618 | 22.644 | 337 | 0.328 | 47.6 | 261 | 0.462 |
| 69 | 23.115 | 328 | 0.616 | 22.977 | 337 | 0.333 | 48.3 | 258 | 0.460 |
| 70 | 23.45 | 328 | 0.614 | 23.31 | 337 | 0.337 | 49 | 254 | 0.456 |
| 71 | 23.785 | 328 | 0.612 | 23.643 | 336 | 0.341 | 49.7 | 250 | 0.451 |
| 72 | 24.12 | 327 | 0.610 | 23.976 | 336 | 0.346 | 50.4 | 246 | 0.444 |
| 73 | 24.455 | 327 | 0.608 | 24.309 | 336 | 0.350 | 51.1 | 241 | 0.436 |
| 74 | 24.79 | 327 | 0.606 | 24.642 | 336 | 0.354 | 51.8 | 235 | 0.427 |
| 75 | 25.125 | 326 | 0.604 | 24.975 | 335 | 0.359 | 52.5 | 228 | 0.417 |
| 76 | 25.46 | 326 | 0.601 | 25.308 | 335 | 0.363 | 53.2 | 219 | 0.406 |
| 77 | 25.795 | 326 | 0.599 | 25.641 | 335 | 0.367 | 53.9 | 392 | 0.393 |
| 78 | 26.13 | 325 | 0.597 | 25.974 | 335 | 0.371 | 54.6 | 385 | 0.380 |
| 79 | 26.465 | 325 | 0.595 | 26.307 | 335 | 0.375 | 55.3 | 379 | 0.366 |
| 80 | 26.8 | 324 | 0.592 | 26.64 | 334 | 0.380 | 56 | 374 | 0.352 |
| 81 | 27.135 | 324 | 0.590 | 26.973 | 334 | 0.384 | 56.7 | 370 | 0.337 |
| 82 | 27.47 | 324 | 0.587 | 27.306 | 334 | 0.388 | 57.4 | 366 | 0.321 |
| 83 | 27.805 | 323 | 0.585 | 27.639 | 334 | 0.392 | 58.1 | 362 | 0.306 |
| 84 | 28.14 | 323 | 0.582 | 27.972 | 333 | 0.396 | 58.8 | 359 | 0.290 |
| 85 | 28.475 | 323 | 0.580 | 28.305 | 333 | 0.400 | 59.5 | 356 | 0.275 |
| 86 | 28.81 | 322 | 0.577 | 28.638 | 333 | 0.404 | 60.2 | 353 | 0.260 |
| 87 | 29.145 | 322 | 0.575 | 28.971 | 333 | 0.408 | 60.9 | 350 | 0.245 |
| 88 | 29.48 | 322 | 0.572 | 29.304 | 333 | 0.412 | 61.6 | 348 | 0.230 |
| 89 | 29.815 | 321 | 0.569 | 29.637 | 332 | 0.416 | 62.3 | 346 | 0.216 |
| 90 | 30.15 | 321 | 0.567 | 29.97 | 332 | 0.420 | 63 | 344 | 0.207 |
| 91 | 30.485 | 321 | 0.564 | 30.303 | 332 | 0.424 | 63.7 | 342 | 0.213 |
| 92 | 30.82 | 320 | 0.561 | 30.636 | 332 | 0.428 | 64.4 | 340 | 0.219 |
| 93 | 31.155 | 320 | 0.558 | 30.969 | 331 | 0.431 | 65.1 | 338 | 0.224 |
| 94 | 31.49 | 320 | 0.556 | 31.302 | 331 | 0.435 | 65.8 | 336 | 0.228 |
| 95 | 31.825 | 319 | 0.553 | 31.635 | 331 | 0.439 | 66.5 | 335 | 0.231 |
| 96 | 32.16 | 319 | 0.550 | 31.968 | 331 | 0.443 | 67.2 | 333 | 0.234 |
| 97 | 32.495 | 319 | 0.547 | 32.301 | 330 | 0.447 | 67.9 | 332 | 0.237 |
| 98 | 32.83 | 318 | 0.544 | 32.634 | 330 | 0.450 | 68.6 | 330 | 0.238 |
| 99 | 33.165 | 318 | 0.541 | 32.967 | 330 | 0.454 | 69.3 | 329 | 0.239 |
| 100 | 33.5 | 318 | 0.539 | 33.3 | 330 | 0.458 | 70 | 327 | 0.240 |
| 101 | 33.835 | 317 | 0.535 | 33.633 | 330 | 0.461 | 70.7 | 326 | 0.240 |
| 102 | 34.17 | 317 | 0.533 | 33.966 | 329 | 0.465 | 71.4 | 325 | 0.240 |
| 103 | 34.505 | 317 | 0.530 | 34.299 | 329 | 0.469 | 72.1 | 324 | 0.239 |
| 104 | 34.84 | 316 | 0.527 | 34.632 | 329 | 0.472 | 72.8 | 323 | 0.237 |
| 105 | 35.175 | 316 | 0.524 | 34.965 | 329 | 0.476 | 73.5 | 321 | 0.236 |

| 2D Material | Graphene | | | h-BN | | | MoS$_2$ | | |
|---|---|---|---|---|---|---|---|---|---|
| Number of Layers | Crystal Thickness $d_1$ (nm) | Vacuum Gap $d_2$ (nm) | Optimal $|\Re|$ (%/nm) | Crystal Thickness $d_1$ (nm) | Vacuum Gap $d_2$ (nm) | Optimal $|\Re|$ (%/nm) | Crystal Thickness $d_1$ (nm) | Vacuum Gap $d_2$ (nm) | Optimal $|\Re|$ (%/nm) |
| 106 | 35.51 | 316 | 0.521 | 35.298 | 328 | 0.479 | 74.2 | 320 | 0.234 |
| 107 | 35.845 | 316 | 0.517 | 35.631 | 328 | 0.483 | 74.9 | 319 | 0.232 |
| 108 | 36.18 | 315 | 0.514 | 35.964 | 328 | 0.486 | 75.6 | 318 | 0.229 |
| 109 | 36.515 | 315 | 0.511 | 36.297 | 328 | 0.490 | 76.3 | 317 | 0.226 |
| 110 | 36.85 | 315 | 0.508 | 36.63 | 328 | 0.493 | 77 | 316 | 0.223 |
| 111 | 37.185 | 314 | 0.505 | 36.963 | 327 | 0.497 | 77.7 | 315 | 0.220 |
| 112 | 37.52 | 314 | 0.502 | 37.296 | 327 | 0.500 | 78.4 | 314 | 0.217 |
| 113 | 37.855 | 314 | 0.499 | 37.629 | 327 | 0.503 | 79.1 | 313 | 0.213 |
| 114 | 38.19 | 313 | 0.496 | 37.962 | 327 | 0.507 | 79.8 | 312 | 0.210 |
| 115 | 38.525 | 313 | 0.492 | 38.295 | 326 | 0.510 | 80.5 | 311 | 0.206 |
| 116 | 38.86 | 313 | 0.489 | 38.628 | 326 | 0.513 | 81.2 | 310 | 0.202 |
| 117 | 39.195 | 312 | 0.486 | 38.961 | 326 | 0.517 | 81.9 | 309 | 0.198 |
| 118 | 39.53 | 312 | 0.483 | 39.294 | 326 | 0.520 | 82.6 | 308 | 0.194 |
| 119 | 39.865 | 312 | 0.479 | 39.627 | 325 | 0.523 | 83.3 | 307 | 0.190 |
| 120 | 40.2 | 311 | 0.476 | 39.96 | 325 | 0.526 | 84 | 306 | 0.185 |
| 121 | 40.535 | 311 | 0.473 | 40.293 | 325 | 0.529 | 84.7 | 305 | 0.181 |
| 122 | 40.87 | 311 | 0.469 | 40.626 | 325 | 0.532 | 85.4 | 304 | 0.177 |
| 123 | 41.205 | 310 | 0.466 | 40.959 | 325 | 0.536 | 86.1 | 303 | 0.172 |
| 124 | 41.54 | 310 | 0.463 | 41.292 | 324 | 0.539 | 86.8 | 302 | 0.167 |
| 125 | 41.875 | 310 | 0.459 | 41.625 | 324 | 0.542 | 87.5 | 301 | 0.163 |
| 126 | 42.21 | 310 | 0.456 | 41.958 | 324 | 0.545 | 88.2 | 300 | 0.158 |
| 127 | 42.545 | 309 | 0.453 | 42.291 | 324 | 0.548 | 88.9 | 299 | 0.154 |
| 128 | 42.88 | 309 | 0.449 | 42.624 | 323 | 0.551 | 89.6 | 298 | 0.149 |
| 129 | 43.215 | 309 | 0.446 | 42.957 | 323 | 0.554 | 90.3 | 296 | 0.144 |
| 130 | 43.55 | 308 | 0.442 | 43.29 | 323 | 0.557 | 91 | 295 | 0.139 |
| 131 | 43.885 | 308 | 0.439 | 43.623 | 323 | 0.559 | 91.7 | 294 | 0.135 |
| 132 | 44.22 | 308 | 0.435 | 43.956 | 323 | 0.562 | 92.4 | 293 | 0.130 |
| 133 | 44.555 | 307 | 0.432 | 44.289 | 322 | 0.565 | 93.1 | 292 | 0.125 |
| 134 | 44.89 | 307 | 0.428 | 44.622 | 322 | 0.568 | 93.8 | 290 | 0.120 |
| 135 | 45.225 | 307 | 0.425 | 44.955 | 322 | 0.571 | 94.5 | 289 | 0.115 |
| 136 | 45.56 | 306 | 0.421 | 45.288 | 322 | 0.573 | 95.2 | 288 | 0.110 |
| 137 | 45.895 | 306 | 0.417 | 45.621 | 321 | 0.576 | 95.9 | 286 | 0.108 |
| 138 | 46.23 | 306 | 0.414 | 45.954 | 321 | 0.579 | 96.6 | 285 | 0.110 |
| 139 | 46.565 | 305 | 0.410 | 46.287 | 321 | 0.582 | 97.3 | 283 | 0.111 |
| 140 | 46.9 | 305 | 0.407 | 46.62 | 321 | 0.584 | 98 | 282 | 0.112 |
| 141 | 47.235 | 305 | 0.403 | 46.953 | 320 | 0.587 | 98.7 | 280 | 0.114 |
| 142 | 47.57 | 304 | 0.399 | 47.286 | 320 | 0.589 | 99.4 | 279 | 0.115 |
| 143 | 47.905 | 304 | 0.396 | 47.619 | 320 | 0.592 | 100.1 | 277 | 0.116 |
| 144 | 48.24 | 304 | 0.392 | 47.952 | 320 | 0.594 | 100.8 | 275 | 0.117 |

| 2D Material | Graphene | | | h-BN | | | MoS$_2$ | | |
|---|---|---|---|---|---|---|---|---|---|
| Number of Layers | Crystal Thickness $d_1$ (nm) | Vacuum Gap $d_2$ (nm) | Optimal $|\Re|$ (%/nm) | Crystal Thickness $d_1$ (nm) | Vacuum Gap $d_2$ (nm) | Optimal $|\Re|$ (%/nm) | Crystal Thickness $d_1$ (nm) | Vacuum Gap $d_2$ (nm) | Optimal $|\Re|$ (%/nm) |
| 145 | 48.575 | 303 | 0.388 | 48.285 | 320 | 0.597 | 101.5 | 273 | 0.118 |
| 146 | 48.91 | 303 | 0.385 | 48.618 | 319 | 0.599 | 102.2 | 271 | 0.118 |
| 147 | 49.245 | 303 | 0.381 | 48.951 | 319 | 0.602 | 102.9 | 269 | 0.118 |
| 148 | 49.58 | 303 | 0.377 | 49.284 | 319 | 0.604 | 103.6 | 266 | 0.119 |
| 149 | 49.915 | 302 | 0.374 | 49.617 | 319 | 0.607 | 104.3 | 264 | 0.118 |
| 150 | 50.25 | 302 | 0.370 | 49.95 | 318 | 0.609 | 105 | 261 | 0.118 |
| 151 | 50.585 | 302 | 0.366 | 50.283 | 318 | 0.611 | 105.7 | 259 | 0.118 |
| 152 | 50.92 | 301 | 0.363 | 50.616 | 318 | 0.614 | 106.4 | 256 | 0.117 |
| 153 | 51.255 | 301 | 0.359 | 50.949 | 318 | 0.616 | 107.1 | 252 | 0.116 |
| 154 | 51.59 | 301 | 0.355 | 51.282 | 317 | 0.618 | 107.8 | 249 | 0.115 |
| 155 | 51.925 | 300 | 0.351 | 51.615 | 317 | 0.621 | 108.5 | 245 | 0.114 |
| 156 | 52.26 | 300 | 0.348 | 51.948 | 317 | 0.623 | 109.2 | 241 | 0.112 |
| 157 | 52.595 | 300 | 0.344 | 52.281 | 317 | 0.625 | 109.9 | 236 | 0.111 |
| 158 | 52.93 | 299 | 0.340 | 52.614 | 317 | 0.627 | 110.6 | 230 | 0.109 |
| 159 | 53.265 | 299 | 0.336 | 52.947 | 316 | 0.629 | 111.3 | 389 | 0.107 |
| 160 | 53.6 | 299 | 0.333 | 53.28 | 316 | 0.631 | 112 | 384 | 0.105 |
| 161 | 53.935 | 298 | 0.329 | 53.613 | 316 | 0.633 | 112.7 | 380 | 0.102 |
| 162 | 54.27 | 298 | 0.325 | 53.946 | 316 | 0.635 | 113.4 | 376 | 0.100 |
| 163 | 54.605 | 298 | 0.321 | 54.279 | 315 | 0.637 | 114.1 | 372 | 0.097 |
| 164 | 54.94 | 297 | 0.318 | 54.612 | 315 | 0.640 | 114.8 | 369 | 0.095 |
| 165 | 55.275 | 297 | 0.314 | 54.945 | 315 | 0.641 | 115.5 | 366 | 0.092 |
| 166 | 55.61 | 297 | 0.310 | 55.278 | 315 | 0.643 | 116.2 | 363 | 0.089 |
| 167 | 55.945 | 296 | 0.306 | 55.611 | 314 | 0.645 | 116.9 | 360 | 0.086 |
| 168 | 56.28 | 296 | 0.303 | 55.944 | 314 | 0.647 | 117.6 | 358 | 0.083 |
| 169 | 56.615 | 296 | 0.299 | 56.277 | 314 | 0.649 | 118.3 | 355 | 0.080 |
| 170 | 56.95 | 295 | 0.295 | 56.61 | 314 | 0.651 | 119 | 353 | 0.077 |
| 171 | 57.285 | 295 | 0.292 | 56.943 | 314 | 0.653 | 119.7 | 351 | 0.074 |
| 172 | 57.62 | 295 | 0.288 | 57.276 | 313 | 0.654 | 120.4 | 349 | 0.070 |
| 173 | 57.955 | 294 | 0.284 | 57.609 | 313 | 0.656 | 121.1 | 347 | 0.067 |
| 174 | 58.29 | 294 | 0.281 | 57.942 | 313 | 0.658 | 121.8 | 345 | 0.064 |
| 175 | 58.625 | 294 | 0.277 | 58.275 | 313 | 0.660 | 122.5 | 343 | 0.061 |
| 176 | 58.96 | 293 | 0.273 | 58.608 | 312 | 0.661 | 123.2 | 342 | 0.058 |
| 177 | 59.295 | 293 | 0.270 | 58.941 | 312 | 0.663 | 123.9 | 340 | 0.055 |
| 178 | 59.63 | 293 | 0.266 | 59.274 | 312 | 0.665 | 124.6 | 339 | 0.055 |
| 179 | 59.965 | 292 | 0.262 | 59.607 | 312 | 0.666 | 125.3 | 337 | 0.056 |
| 180 | 60.3 | 292 | 0.259 | 59.94 | 311 | 0.668 | 126 | 336 | 0.057 |
| 181 | 60.635 | 291 | 0.255 | 60.273 | 311 | 0.669 | 126.7 | 334 | 0.057 |
| 182 | 60.97 | 291 | 0.251 | 60.606 | 311 | 0.671 | 127.4 | 333 | 0.058 |
| 183 | 61.305 | 291 | 0.248 | 60.939 | 311 | 0.672 | 128.1 | 331 | 0.059 |

| 2D Material | Graphene | | | h-BN | | | MoS$_2$ | | |
|---|---|---|---|---|---|---|---|---|---|
| Number of Layers | Crystal Thickness $d_1$ (nm) | Vacuum Gap $d_2$ (nm) | Optimal $|\Re|$ (%/nm) | Crystal Thickness $d_1$ (nm) | Vacuum Gap $d_2$ (nm) | Optimal $|\Re|$ (%/nm) | Crystal Thickness $d_1$ (nm) | Vacuum Gap $d_2$ (nm) | Optimal $|\Re|$ (%/nm) |
| 184 | 61.64 | 290 | 0.244 | 61.272 | 310 | 0.674 | 128.8 | 330 | 0.059 |
| 185 | 61.975 | 290 | 0.241 | 61.605 | 310 | 0.675 | 129.5 | 329 | 0.060 |
| 186 | 62.31 | 290 | 0.237 | 61.938 | 310 | 0.677 | 130.2 | 328 | 0.060 |
| 187 | 62.645 | 289 | 0.234 | 62.271 | 310 | 0.678 | 130.9 | 326 | 0.060 |
| 188 | 62.98 | 289 | 0.230 | 62.604 | 310 | 0.679 | 131.6 | 325 | 0.061 |
| 189 | 63.315 | 289 | 0.227 | 62.937 | 309 | 0.681 | 132.3 | 324 | 0.061 |
| 190 | 63.65 | 288 | 0.223 | 63.27 | 309 | 0.682 | 133 | 323 | 0.060 |
| 191 | 63.985 | 288 | 0.220 | 63.603 | 309 | 0.683 | 133.7 | 322 | 0.060 |
| 192 | 64.32 | 288 | 0.216 | 63.936 | 309 | 0.684 | 134.4 | 321 | 0.060 |
| 193 | 64.655 | 287 | 0.213 | 64.269 | 308 | 0.686 | 135.1 | 319 | 0.060 |
| 194 | 64.99 | 287 | 0.210 | 64.602 | 308 | 0.687 | 135.8 | 318 | 0.059 |
| 195 | 65.325 | 286 | 0.206 | 64.935 | 308 | 0.688 | 136.5 | 317 | 0.059 |
| 196 | 65.66 | 286 | 0.203 | 65.268 | 308 | 0.689 | 137.2 | 316 | 0.058 |
| 197 | 65.995 | 286 | 0.200 | 65.601 | 307 | 0.690 | 137.9 | 315 | 0.057 |
| 198 | 66.33 | 285 | 0.196 | 65.934 | 307 | 0.691 | 138.6 | 314 | 0.057 |
| 199 | 66.665 | 285 | 0.193 | 66.267 | 307 | 0.693 | 139.3 | 313 | 0.056 |
| 200 | 67 | 285 | 0.190 | 66.6 | 307 | 0.694 | 140 | 312 | 0.055 |